\documentclass[aps,prl,twocolumn,showpacs, superscriptaddress, floatfix]{revtex4}
\usepackage{graphicx}
\usepackage{color}
\usepackage{float}
\usepackage{rotating}

\begin{document}
\title{Photo-induced stabilization and enhancement of the ferroelectric polarization in Ba$_{0.1}$Sr$_{0.9}$TiO$_{3}$/La$_{0.7}$Ca(Sr)$_{0.3}$MnO$_{3}$ thin film heterostructures}

\author{Y. M. Sheu}
\affiliation{Center for Integrated Nanotechnologies, MS K771, Los Alamos National Laboratory, Los Alamos, New Mexico 87545, USA}
\author{S. A. Trugman}
\affiliation{Center for Integrated Nanotechnologies, MS K771, Los Alamos National Laboratory, Los Alamos, New Mexico 87545, USA}
\author{L. Yan}
\affiliation{Center for Integrated Nanotechnologies, MS K771, Los Alamos National Laboratory, Los Alamos, New Mexico 87545, USA}
\author{C.-P. Chuu}
\affiliation{Institute of Atomic and Molecular Sciences, Academia Sinica, Taipei 10617, Taiwan}
\author{Z. Bi}
\affiliation{Center for Integrated Nanotechnologies, MS K771, Los Alamos National Laboratory, Los Alamos, New Mexico 87545, USA}
\author{Q. X. Jia}
\affiliation{Center for Integrated Nanotechnologies, MS K771, Los Alamos National Laboratory, Los Alamos, New Mexico 87545, USA}
\author{A. J. Taylor}
\affiliation{Center for Integrated Nanotechnologies, MS K771, Los Alamos National Laboratory, Los Alamos, New Mexico 87545, USA}
\author{R. P. Prasankumar}
\affiliation{Center for Integrated Nanotechnologies, MS K771, Los Alamos National Laboratory, Los Alamos, New Mexico 87545, USA}

\date{\today}

\begin{abstract}

An emerging area in condensed matter physics is the use of multilayered heterostructures to enhance ferroelectricity in complex oxides. Here, we demonstrate that optically pumping carriers across the interface between thin films of a ferroelectric (FE) insulator and a ferromagnetic metal can significantly enhance the FE polarization. The photoinduced FE state remains stable at low temperatures for over one day. This occurs through screening of the internal electric field by the photoexcited carriers, leading to a larger, more stable polarization state that may be suitable for applications in areas such as data and energy storage.

\end{abstract}
 \maketitle

\begin{figure}[H]
\begin{center}
\includegraphics[width=3.5in]{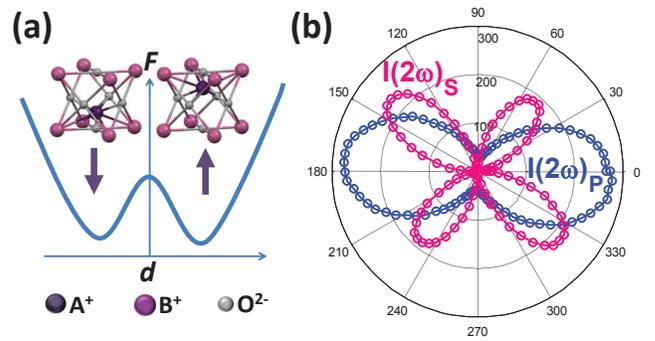}
\caption{ \label{f:FE} (color online)  (a) DWP energy ($F$) as a function of A$^{+}$ displacement ($d$) in the FE tetragonal structure. The A$^{+}$ (Ti$^{4+}$ in BSTO) ion displaces from the center, breaking inversion symmetry along the tetragonal axis. (b) Polar plot of the SHG signal in BSTO/LCMO/MgO detected for both $s$ and $p$ polarizations. The SHG signal originates from the C$_{4v}$ symmetry, with the optical axis along the tetragonal long axis.  }
\end{center}
\end{figure}

The wide variety of applications for ferroelectric (FE) materials \cite{Scott2007Science} has motivated many fundamental studies aimed at controlling their properties, particularly by increasing the FE polarization and its transition temperature T$_{c}$. Arguably the most heavily studied class of ferroelectrics are the highly polarizable perovskite oxides ABO$_3$ (e.g., BaTiO$_3$), in which the center cation (A$^{+}$) is displaced along one direction with respect to the surrounding oxygen octahedron (O$^{2-}$), resulting in a spontaneous polarization that can be switched between different potential wells with an applied external electric (E) field below T$_{c}$ (Fig. \ref{f:FE}(a)). Interest in these materials substantially increased when they were first fabricated in thin film form in the 1980s, since this enabled their integration into semiconductor chips \cite{Scott2007Science}.

Strain can provide an avenue for tuning the properties of thin FE perovskite oxide films by forcing their in-plane lattice constant ($a_{in}$) to match that of the substrate \cite{Choi2004Science}. This reduces the number of possible displacement directions (i.e., positive and negative directions along the edges or diagonals of a cube) for the A$^{+}$ ion. Therefore, even in the paraelectric (PE) phase the symmetry is lowered \cite{Pertsev1998PRL}, resulting in an increase in T$_{c}$ \cite{Choi2004Science,Pertsev1998PRL,Shirokov2009PRB}. For example, in the case of compressive strain on a cubic substrate, elongation along the surface normal reduces the number of possible displacement directions to two (either up or down along the surface normal). The free energy of this system can be represented by a double-well potential (DWP) (Fig. \ref{f:FE}(a)), in which the lower potential well is determined by boundary conditions (e.g. strain, E field, interface atomic structures, etc.) that sets the direction of A$^+$ displacement and thus the FE polarization.  However, this displacement is reduced when the FE polarization is terminated at the film interfaces, as this creates a large internal "depolarizing" E ($E_{\text{{int}}}$) field in the opposite direction that reduces remanent polarization (P$_{r}$, the polarization without an external applied E field) \cite{Scott2007Science,Junquera2003Nature,Batra1973PRB}. This can be overcome by adding charge at the interfaces to screen the depolarizing field, allowing P$_{r}$ to nearly reach the bulk value \cite{Junquera2003Nature}.

This fundamental coupling between $E_{\text{{int}}}$ and P$_{r}$ limits a variety of nanoscale device applications (e.g., data storage), since various complex boundary conditions (e.g., interface electrostatic structure and chemical bonding) usually result in incomplete charge screening that is sensitive to external parameters \cite{Junquera2003Nature}. Therefore, a substantial portion of current research focuses on obtaining high P$_{r}$ at the nanoscale \cite{Junquera2003Nature,Choi2004Science,Lee2005Nature,Spanier2006NanoLett}.

In this context, it is surprising that relatively little work has explored the use of optical methods for enhancing P$_{r}$, particularly when an FE film is combined with another complex oxide film in a multilayered structure to provide additional functionality. One of the most straightforward ways to optically enhance ferroelectricity would be to photoexcite carriers that screen the depolarizing E field. However, in individual FE films, electron-hole recombination limits the carrier lifetime, minimizing this enhancement  \cite{Sheu2012APL}. This could be overcome by using a bilayered heterostructure, in which one species of photoexcited carrier (e.g., holes) is isolated on the non-FE side of the interface and the other species (e.g., electrons) is isolated in the FE film, minimizing recombination. This would result in physically separated, long-lived screening charges, potentially leading to high P$_{r}$.

In this Letter, we demonstrate a non-contact, all-optical method for writing and reading the FE polarization in Ba$_{0.1}$Sr$_{0.9}$TiO$_{3}$ (BSTO) thin films that are grown on ferromagnetic (FM) metallic manganite thin films on different substrates. The lattice mismatch between BSTO and the manganite layers fully strains the BSTO films, giving rise to higher FE T$_c$'s than in the bulk. More importantly, by photoexciting the manganite films at 1.59 eV, we were able to enhance P$_r$ through charge transfer across the BSTO/manganite interface, as revealed by measuring the resulting 3.18 eV second-harmonic generation (SHG) signal from the FE polarization in BSTO. We observe that this new state has a very long lifetime (over one day) after removal of the initial photoexcitation, suggesting that growth of FE/manganite bilayers offers a new avenue for optically increasing and stabilizing FE order.

Our SHG experiments are based on an amplified Ti:sapphire laser system producing pulses at a 250 kHz repetition rate with $\sim$100 fs duration and energies of $\sim$4 $\mu$J/pulse at a center wavelength of 780 nm (1.59 eV). Using a single laser beam, SHG is generated from BSTO at 3.18 eV upon photoexciting the manganite films at 1.59 eV, after which it is detected by a photomultiplier tube (PMT) using lock-in detection after filtering out the fundamental signal. The fundamental light polarization is controlled by a half-wave plate and the SHG signal is detected for either \textit{p} or \textit{s} polarizations. The standard laser fluence used for SHG in our measurements is $F_0\sim$0.25 mJ/cm$^{2}$.

The samples used in our studies are 50 nm thick BSTO films grown by pulsed laser deposition on 50 nm thick optimally doped manganite films (La$_{0.7}$Ca$_{0.3}$MnO$_{3}$ (LCMO) or La$_{0.7}$Sr$_{0.3}$MnO$_{3}$ (LSMO)), using MgO or SrTiO$_{3}$ (STO) (001) substrates. The substrate temperature during film growth is initially optimized and maintained at 750  $^{o}$C. The oxygen pressure during deposition is 100 mTorr. The samples are cooled to room temperature in pure oxygen (350 Torr) without further thermal treatment.

The structural properties of our samples were then determined by X-ray diffraction (XRD) (Table 1). In the 10$\%$ Ba-doped STO films studied here, previous work has revealed a rich phase diagram when varying the strain \cite{Shirokov2009PRB}. In particular, above 0.2$\%$ in-plane compressive strain, elongation along the surface normal results in a transition to a tetragonal structure below 75 K, in which the FE polarization either points up or down along the $c$ axis (associated with tetragonal $C_{4v}$ symmetry) (Fig. \ref{f:FE}(a)).  In our samples, XRD shows that the small bulk lattice mismatch between LC(S)MO and STO causes the LC(S)MO films to be fully strained, matching the $a_{in}$ of STO. The mismatch between LC(S)MO and MgO is much larger, so that the LC(S)MO films are relaxed or partially relaxed. Furthermore, all BSTO layers are fully compressively strained to match the $a_{in}$ of LC(S)MO (Table 1) and thus have a tetragonal structure at room temperature. This is far above previously measured FE T$_{c}$'s for single crystals or thick films, which are typically $\sim$70-80 K \cite{Lemanov1996PRB,Shirokov2009PRB,Tenne2004PRB}. However, this does not necessitate the existence of a room temperature FE phase in our films, as discussed in more detail below \cite{Choi2004Science,Pertsev1998PRL}.

\begin{figure}[tb]
\begin{center}
\includegraphics[width=3.45in]{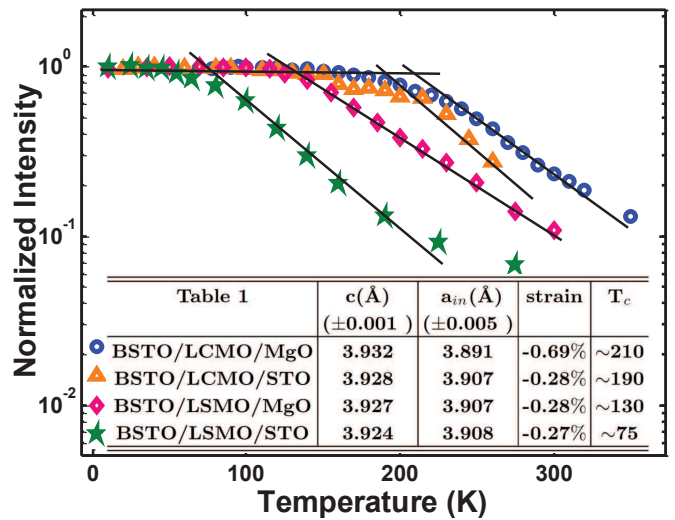}
\caption{ \label{f:Tc} (color online) SHG intensity of various films measured for ($p_{in}$,$p_{out}$). Each data set is normalized to its individual signal at 10 K. Inset table lists the out-of-plane (c), in-plane (a$_{in}$) lattice constant, the in-plane strain, and T$_{c}$ of BSTO on various LC(S)MO/substrate, calculated from XRD.}
\end{center}
\end{figure}

In general, SHG  originates from the high field-induced polarization P$_{i}(2\omega)=\chi^{(2)}_{ijk}$E$_{j}(\omega)$E$_{k}(\omega)$, where E$_{j}$ and E$_{k}$ are the incident electric field of the fundamental light and $\chi^{(2)}_{ijk}$ is the tensor describing the second order nonlinear susceptibility, which only has a non-zero component when the crystal inversion symmetry is broken \cite{SHGBook}. In the PE phase, the strained BSTO film has a tetragonal structure (with $C_4$ symmetry), but there is no Ti$^{4+}$ displacement to break inversion symmetry, and thus no significant SHG signal. In the FE phase, Ti$^{4+}$ displacement breaks inversion symmetry, leading to an FE polarization directed along the surface normal ($C_{4v}$ symmetry). In this symmetry, SHG is only allowed when the fundamental E field has a component along the FE dipole, thus exhibiting a sinusoidal dependence on the incident angle (as observed in separate angle-dependent measurements on our samples). Therefore, we used a reflection geometry, in which the fundamental light was incident at a $\sim$23$^{\text{o}}$ angle, to obtain the data shown here. Fig. \ref{f:FE}(b) depicts the \textit{s} and \textit{p} polarized SHG signals measured as a function of incident light polarization. We analyze the observed pattern to confirm the $C_{4v}$ symmetry \cite{Choi2004Science}. We also performed experiments on individual manganite films to verify that the LC(S)MO layers do not contribute to the measured SHG signal.

Fig. \ref{f:Tc} displays the temperature dependence of the SHG signal for all four of the samples studied here with both fundamental and SHG light \textit{p} polarized (abbreviated $p_{in}$, $p_{out}$), which is the configuration that consistently gives the largest SHG signal for all samples. The intersection of the two different slopes in the data indicates T$_{c}$ \cite{Choi2004Science}, below which the tetragonal $c$ axis contracts when heating. We find that T$_{c}$ is highest in the BSTO/LCMO/MgO sample, consistent with the fact that the BSTO layer is under the highest strain. The variation in T$_{c}$ for the other samples with comparable strain values is likely due to the complexity of the phase diagram for BSTO in this strain range, particularly for low Ba doping \cite{Shirokov2009PRB}. This makes T$_{c}$ very sensitive to other factors, such as the high photoinduced carrier concentrations discussed below.

\begin{figure}[tb]
\begin{center}
\includegraphics[width=3.0in]{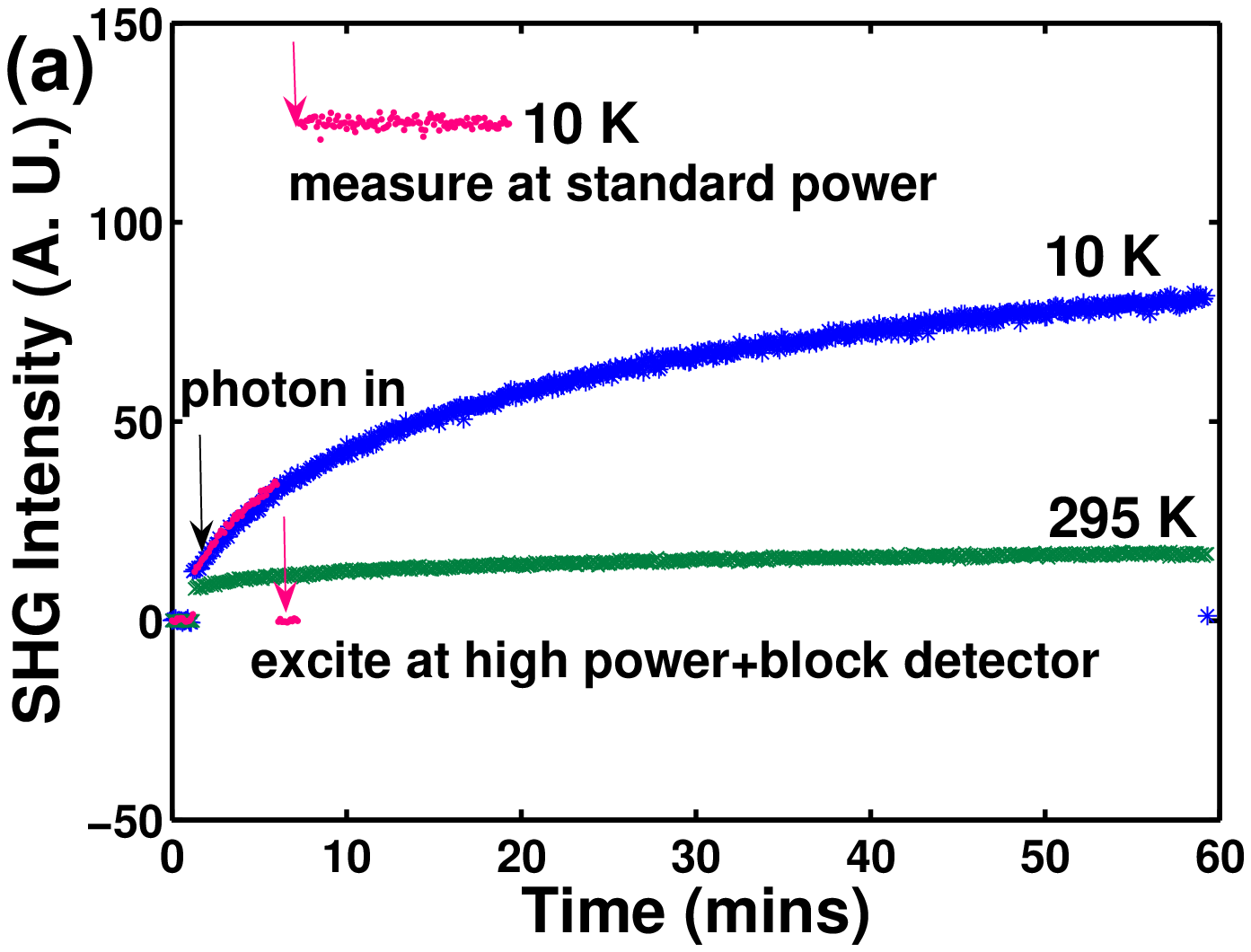}
\includegraphics[width=3.2in]{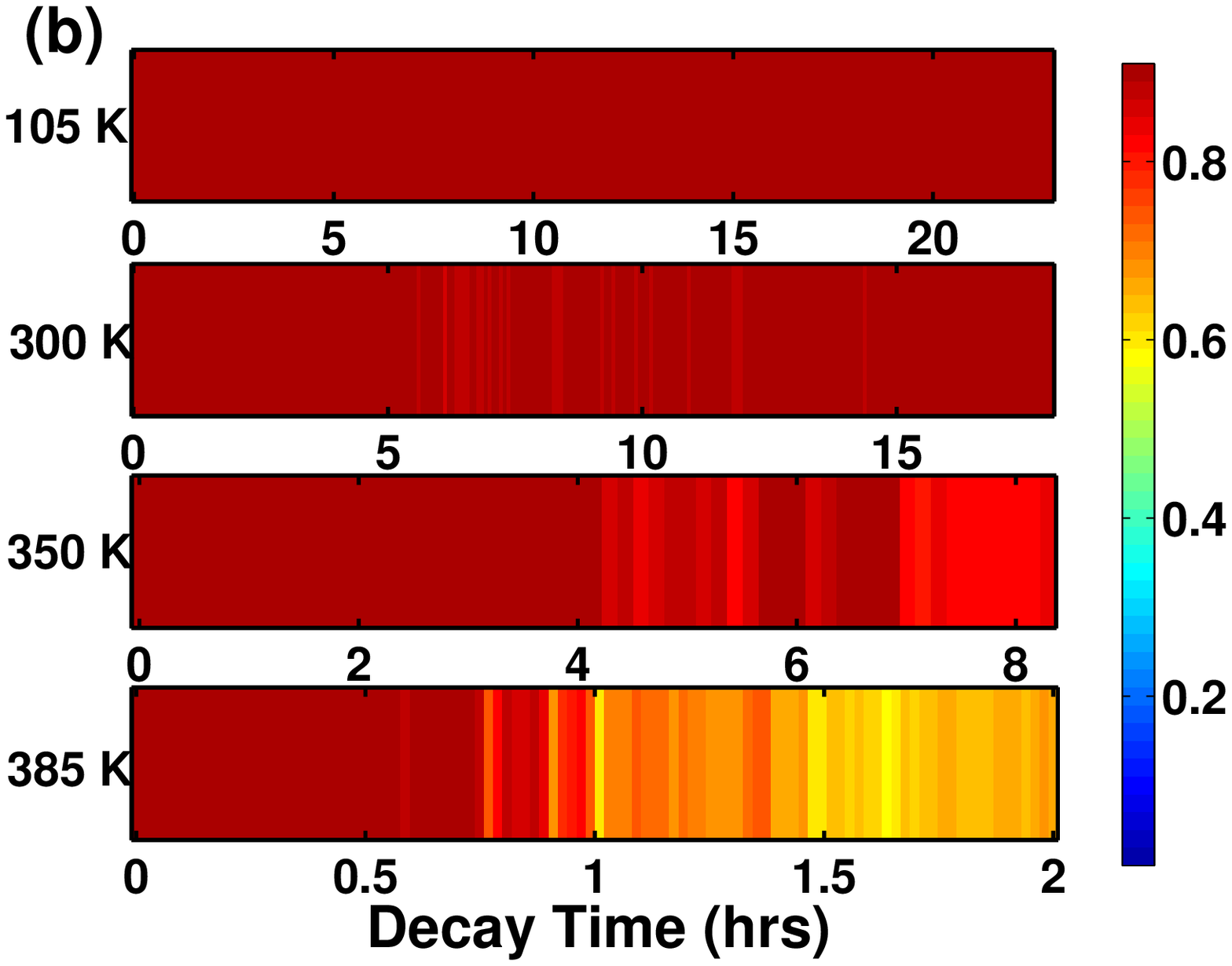}
\caption{ \label{f:saturation_decay} (color online) (a) The time-varying SHG signals in BSTO/LSMO/MgO at low and high temperatures. The detailed measurement process and the difference between traces are explained in the text. (b) Decay of the SHG signal measured at various temperatures after saturation upon removal of photoexcitation. There is no observable decay within a day at low temperatures. }
\end{center}
\end{figure}

Fig. \ref{f:saturation_decay}(a) shows the main result of this work, which is our striking observation of a slow increase in the SHG signal that nearly saturates after tens of minutes.  To generate the data in Fig. \ref{f:saturation_decay}(a), we started at a fresh spot on the sample, recorded a "dark" signal with no laser exposure for the first minute to calibrate the background level, and then exposed the sample to the laser beam at $F_0\sim$0.25 mJ/cm$^{2}$ to generate a SHG signal. The trace depicting a discrete jump to the saturated state at 10 K (pink curve) was obtained by exposing the sample to a fluence nearly one order of magnitude larger while blocking the detector for one minute, so that the SHG signal fell to zero. After this treatment, we again measured the SHG signal at $F_0$, revealing that it immediately jumped to the saturated value, bypassing the slow increase observed for lower intensities, after which it remained nearly constant. In contrast, the data in Fig. \ref{f:Tc} was taken while heating the sample, after exposing a fixed spot on the sample to the laser beam at 10 K until the SHG signal was nearly saturated; this procedure ensured that all of the temperature-dependent data was taken in the saturated state.

\begin{figure*}[tb]
\begin{center}
\includegraphics[width=7in]{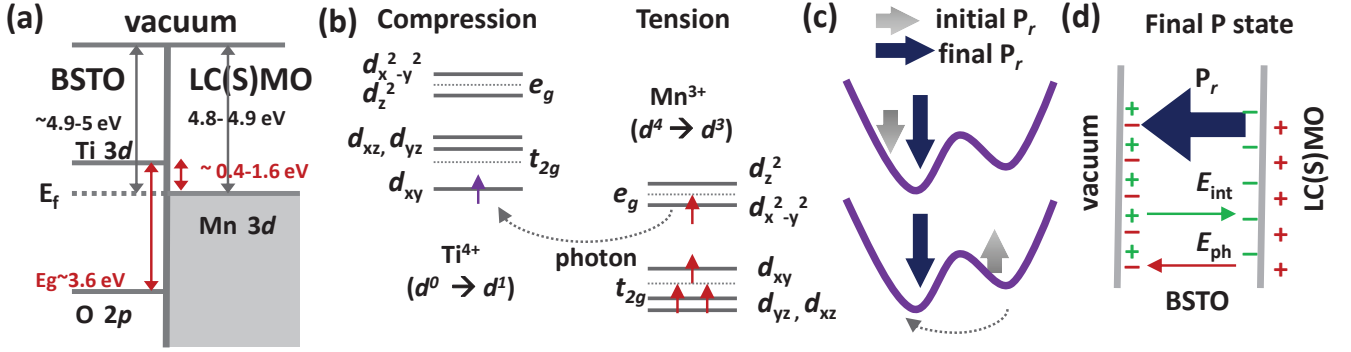}
\caption{ \label{f:Screening} (color online) (a) Schematic band diagram of the BSTO/LC(S)MO interface \cite{GapNote,BandNote}. (b) Diagram of (tetragonal) crystal field 3$d$ orbital occupation in Ti$^{4+}$ and Mn$^{3+}$ under compressive and tensile strain, respectively. The dotted lines depict degenerate levels before they are split by the crystal distortion. Before photoexcitation, Ti$^{4+}$ has an empty $d$ orbital while Mn$^{3+}$ has 3 electrons that half fill the $t_{2g}$ orbital and 1 electron that occupies an $e_{g}$ orbital \cite{MagneticOxide}. Optical excitation promotes one electron across the interface from Mn$^{3+}$ to Ti$^{4+}$, stabilizing the J-T distortion. (c) The final P$_r$ is set by photo-carriers that unbalance the DWP energy in the same way, favoring one polar direction irrespective of the initial direction (see text for more detail). (d) After photoexcitation, $e_{\text{ph}}$ propagate towards the BSTO surface, leaving holes behind at the LC(S)MO interface. Green charges represent the initial FE polarization, and red charges represent photoexcited carriers (creating $E_{\text{{ph}}}$) that screen $E_{\text{{int}}}$, enhancing P$_{r}$ and configuring the final polarization to point towards the BSTO surface (see text for more detail). }
\end{center}
\end{figure*}

The observed saturation phenomenon has the following main features in common for all of our BSTO films: (a) As the laser beam moves to a new spot, the entire process repeats itself, with the SHG signal growing from low to high values for all BSTO/manganite bilayers.  (b) The saturated signal is much larger at low temperatures than at high temperatures (Fig. \ref{f:saturation_decay}(a)), especially below T$_{c}$ in each film. (c) SHG saturation is suppressed in pure BSTO films grown directly on STO or MgO substrates. (d) The timescale required to reach saturation is different at each temperature. (e) The long time required to reach saturation at $F_0$ can be circumvented when a much stronger fluence is used before the sample is saturated (pink curve in Fig. \ref{f:saturation_decay}(a)). When comparing signals from samples that have undergone this treatment, the SHG is always detected at $F_0$. (f) The SHG signal has a strong correlation with T$_{c}$: the higher T$_{c}$, the higher the SHG signal, both initially and in the saturated state. (g) The SHG symmetry (Fig. \ref{f:FE}(b)) does not change with time; only the intensity varies with time. Saturation of the SHG signal is observed in any configuration of fundamental and SHG polarizations, as long as there is a detectable SHG signal. (h) The saturation is suppressed and the high SHG signal disappears when the sample is exposed to air, both of which recur when the sample is returned to vacuum.

Another noteworthy result from our measurements is the extremely long decay time of the photoinduced saturated state. There is no observable decay for over one day at low temperatures (upper panel of Fig. \ref{f:saturation_decay}(b)) after saturation is reached and the laser beam is blocked (only unblocking it to sample the SHG signal). After a day with no discernible decay we moved the beam to a fresh spot, after which it repeated the saturation process. When we moved the beam back to the original spot, the SHG signal returned with the same intensity. The SHG signal only decays at higher temperatures ($>$300 K), with a time constant of $\sim$3 hours at 385 K (lower panel of Fig. \ref{f:saturation_decay}(b)) that decreases with increasing temperature.

All of the above phenomena are robustly reproducible. Similar phenomena were previously studied in Si/SiO$_{2}$ \cite{Bloch1996PRLSHG,Mihaychuk1995OLSHG,Lu2008PRBSHGT}. However, except for the observation of SHG saturation in itself, those results conflict with our data, including the dependence on temperature, power, and symmetry, as well as the decay time.

To explain our observations, we first note that this complex time-varying phenomenon is likely related to photoinduced charge transport across the interface between BSTO and LC(S)MO, since it is suppressed in a pure BSTO film and LC(S)MO alone does not contribute to our SHG signal. We can support this hypothesis by estimating the photoinduced interface charge density, $\sim5\times$10$^{13}$ cm$^{-2}$, assuming that 10$\%$ of the pump light is absorbed in LC(S)MO near the interface. This corresponds to a polarization (P=$Q/A$, where $Q$ is the charge and $A$ is the area of the laser spot) of $\sim$8 $\mu$C/cm$^{2}$ in the saturated state. The overall $\sim$500-800$\%$ increase in SHG intensity from the initial state to the saturated state at 10 K (Fig. \ref{f:saturation_decay}(a)) corresponds to a $\sim$200-300 $\%$ enhancement in the FE polarization, indicating that the initial P$_{r}$$\sim$2-4 $\mu$C/cm$^{2}$.   This is comparable to previously reported values for P$_{r}$ both near this Ba concentration  \cite{Lemanov1996PRB} and at higher concentrations \cite{Tinte2004Pr,Adikary2003Pr,Pontes2001Pr,Ruckenbauer2004Pr,Lee2012NLPr},  validating our hypothesis that the polarization enhancement is linked to the photoinduced interface charge density.

We can draw additional conclusions from the other observations described above. For example, the long-lived nature of the photoinduced state suggests that recombination of the photoexcited carriers is negligible. Furthermore, the strong effect of air exposure on the observed phenomena indicates involvement of the BSTO surface charge. In addition, ferromagnetism in the LC(S)MO layers does not appear to play a role in the observed phenomena on the timescales discussed here. Finally, the lack of variation in the SHG polar pattern with time demonstrates that the symmetry of BSTO remains the same upon photoexcitation, indicating that the observed slow increase in the SHG signal is due to further displacement of the Ti$^{4+}$ ions along the tetragonal axis.

Overall, these considerations suggest that the photo-induced enhancement of the FE polarization is associated with charge screening at the BSTO surface and BSTO/LC(S)MO interface that compensates the $E_{\text{{int}}}$ originating from FE dipoles, making further displacement of Ti$^{4+}$ energetically favorable. We note that expansion of BSTO along its tetragonal axis could also lead to an increase in the Ti$^{4+}$ displacement and consequent increase in the SHG signal. However, photoinduced heating in LC(S)MO causes all three of its crystal axes to expand; the expansion in the $ab$ plane of LC(S)MO will cause the $c$ axis of BSTO to contract, and thus cannot explain our observations.

Based on the above considerations, we provide a phenomenological model to explain the mechanism, neglecting more complex interfacial effects such as band bending and hybridization between Mn and Ti atoms.  We begin by microscopically examining the photoinduced charge transfer across the interface. Fig. \ref{f:Screening}(a) displays a schematic band diagram of the interface between insulating BSTO and metallic LC(S)MO. Conduction in LC(S)MO originates from Mn(3$d$) electrons, while the gap in BSTO is between valence O(2$p$) states and conduction Ti(3$d$) states (with a bandgap of $\sim$3.3-3.9 eV \cite{CardonaPR1965,Thielsch1997Gap,Tian2001Gap,Samantaray2004Gap,Pontes2002Gap,GapNote}). Our XRD data indicates that the Ti$^{4+}$ and Mn$^{3+}$ ions are under compressive and tensile strain, respectively, causing the doublet $e_{g}$ and triplet $t_{2g}$ levels associated with both ions to split \cite{MagneticOxide}. In the FE phase of BSTO the resulting elongation of the $c$ axis splits the Ti $t_{2g}$ states into a doublet ($d_{z^2}$, $d_{x^2-y^2}$) and a singlet $d_{xy}$, stabilizing the Jahn-Teller (J-T) distortion, which is specifically sensitive to the strain \cite{MagneticOxide} (Fig. \ref{f:Screening}(b)). Photoexciting the system at 1.59 eV, which is smaller than the band gap of BSTO but larger than the band offset between the BSTO conduction band and the LC(S)MO Fermi level \cite{BandNote} (Fig. \ref{f:Screening}(a)), transfers one electron from LC(S)MO across the interface to occupy the Ti$^{4+}$ $d_{xy}$ level, changing its valence to Ti$^{3+}$ and further favoring the J-T distortion.

If these photoexcited electrons ($e_{\text{ph}}$) remain near the interface, they will quickly recombine with the photoexcited holes ($h_{\text{ph}}$).  This can be avoided if the static FE polarization points towards the BSTO surface, as $E_{\text{{int}}}$ (pointing in the opposite direction) will cause $e_{\text{ph}}$ to move towards the BSTO surface, while $h_{\text{ph}}$ remain in LC(S)MO near the interface. These photoexcited charges will then compensate the static FE dipoles at both interfaces, reducing $E_{\text{{int}}}$ and enhancing the FE polarization (Fig. \ref{f:Screening}(d)).

In fact, even if the static FE polarization initially points towards the BSTO/LC(S)MO interface, we expect that the final P$_{r}$ will point towards the BSTO surface. This can be seen by considering the photoinduced changes in the DWP. The separation of $e_{\text{ph}}$ and $h_{\text{ph}}$ across the interface produces a large interfacial electric field ($E_{\text{ph}} \gg E_{\text{{int}}}$ ) directed towards the BSTO surface, which acts like an applied E field
that unbalances the DWP energy (Fig. \ref{f:Screening}(c)) and displaces Ti$^{4+}$ ions towards the surface, causing P$_{r}$ to point in that direction.

Further support for this mechanism comes from noting that holes in LC(S)MO cannot cross the interface, as this would require one electron to move from BSTO to LC(S)MO, which is not possible since the Ti$^{4+}$ $d$ orbital is empty and the BSTO O$(2p)$ states are too far away in energy. If $h_{\text{ph}}$ move away from the interface into LC(S)MO, then the energy required to sustain their separation from the $e_{\text{ph}}$ is greater than that gained through compensation of the BSTO surface charge. Photoexcited holes can thus minimize their energy by remaining at the interface.

Our results also suggest that the higher T$_{c}$'s measured in our samples (Fig. \ref{f:Tc}) as compared to previous work in which the strain was varied  \cite{Shirokov2009PRB} is associated with the screening of $E_{\text{{int}}}$ by photoexcited carriers. Similarly higher T$_{c}$'s were previously observed in BaTiO$_{3}$ films sandwiched between metallic electrodes \cite{Choi2004Science}, even though the magnitude of the strain was similar to that in a single film. The reason for the higher T$_{c}$'s observed in that work was not given; however, our work indicates that it likely originates from screening by carriers in the metal electrodes \cite{Batra1973PRB}.

Finally, the long-lived FE polarization observed here at low temperatures is consistent with electrostatic force microscopy (EFM) measurements \cite{Spanier2006NanoLett}. In that work, once the FE polarization was written by applying a voltage, it remained relatively stable for days and only became unstable above T$_{c}$. We did not use this method to determine the T$_{c}$ of our samples, since even at high temperatures the decay is still quite slow, which could lead to overestimation of the T$_{c}$. However, when examining two samples at 350 K, the sample with lower T$_{c}$ (BSTO/LSMO/STO) decays more rapidly than the one with higher T$_{c}$ (BSTO/LSMO/MgO), as expected.

In summary, using an optical-write SHG-read technique, we are able to create an enhanced polarization state that remains stable for over one day in BSTO/LC(S)MO heterostructures. The magnitude of the estimated initial and final remanent polarization agrees with previously reported values. The "ON" and "OFF" processes initiated through photoexcitation in vacuum and through exposure to air, respectively, opens the possibility of non-contact optically-controlled data storage. In addition, the long-lived photoinduced FE state stores photoexcited carriers by confining one species at the interface and another at the surface, making it a good candidate for solar energy storage \cite{Heber2009Nature,Huang2010NP} below T$_{c}$. This work thus represents an excellent example of the novel functionality that can result from combining different complex oxides.

This work was performed at the Center for Integrated Nanotechnologies, a U.S. Department of Energy, Office of Basic Energy Sciences user facility and under the auspices of the Department of Energy, Office of Basic Energy Sciences, Division of Material Sciences. It was also partially supported by the NNSA's Laboratory Directed Research and Development Program. Los Alamos National Laboratory, an affirmative action equal opportunity employer, is operated by Los Alamos National Security, LLC, for the National Nuclear Security Administration of the U. S. Department of Energy under contract DE-AC52-06NA25396.


\begin{thebibliography}{33}
\expandafter\ifx\csname natexlab\endcsname\relax\def\natexlab#1{#1}\fi
\expandafter\ifx\csname bibnamefont\endcsname\relax
  \def\bibnamefont#1{#1}\fi
\expandafter\ifx\csname bibfnamefont\endcsname\relax
  \def\bibfnamefont#1{#1}\fi
\expandafter\ifx\csname citenamefont\endcsname\relax
  \def\citenamefont#1{#1}\fi
\expandafter\ifx\csname url\endcsname\relax
  \def\url#1{\texttt{#1}}\fi
\expandafter\ifx\csname urlprefix\endcsname\relax\def\urlprefix{URL }\fi
\providecommand{\bibinfo}[2]{#2}
\providecommand{\eprint}[2][]{\url{#2}}

\bibitem[{\citenamefont{Scott}(2007)}]{Scott2007Science}
\bibinfo{author}{\bibfnamefont{J.~F.} \bibnamefont{Scott}},
  \bibinfo{journal}{Science} \textbf{\bibinfo{volume}{315}},
  \bibinfo{pages}{954} (\bibinfo{year}{2007}).

\bibitem[{\citenamefont{Choi et~al.}(2004)}]{Choi2004Science}
\bibinfo{author}{\bibfnamefont{K.~J.} \bibnamefont{Choi}} \bibnamefont{et~al.},
  \bibinfo{journal}{Science} \textbf{\bibinfo{volume}{306}},
  \bibinfo{pages}{1005} (\bibinfo{year}{2004}).

\bibitem[{\citenamefont{Pertsev et~al.}(1998)}]{Pertsev1998PRL}
\bibinfo{author}{\bibfnamefont{N.~A.} \bibnamefont{Pertsev}}
  \bibnamefont{et~al.}, \bibinfo{journal}{Phys. Rev. Lett.}
  \textbf{\bibinfo{volume}{80}}, \bibinfo{pages}{1988} (\bibinfo{year}{1998}).

\bibitem[{\citenamefont{Shirokov et~al.}(2009)}]{Shirokov2009PRB}
\bibinfo{author}{\bibfnamefont{V.~B.} \bibnamefont{Shirokov}}
  \bibnamefont{et~al.}, \bibinfo{journal}{Phys. Rev. B}
  \textbf{\bibinfo{volume}{79}}, \bibinfo{pages}{144118}
  (\bibinfo{year}{2009}).

\bibitem[{\citenamefont{Junquera and Ghosez}(2003)}]{Junquera2003Nature}
\bibinfo{author}{\bibfnamefont{J.}~\bibnamefont{Junquera}} \bibnamefont{and}
  \bibinfo{author}{\bibfnamefont{P.}~\bibnamefont{Ghosez}},
  \bibinfo{journal}{Nature} \textbf{\bibinfo{volume}{422}},
  \bibinfo{pages}{506} (\bibinfo{year}{2003}).

\bibitem[{\citenamefont{Batra et~al.}(1973)}]{Batra1973PRB}
\bibinfo{author}{\bibfnamefont{I.~P.} \bibnamefont{Batra}}
  \bibnamefont{et~al.}, \bibinfo{journal}{Phys. Rev. B}
  \textbf{\bibinfo{volume}{8}}, \bibinfo{pages}{3257} (\bibinfo{year}{1973}).

\bibitem[{\citenamefont{Lee et~al.}(2005)}]{Lee2005Nature}
\bibinfo{author}{\bibfnamefont{H.~N.} \bibnamefont{Lee}} \bibnamefont{et~al.},
  \bibinfo{journal}{Nature} \textbf{\bibinfo{volume}{433}},
  \bibinfo{pages}{395} (\bibinfo{year}{2005}).

\bibitem[{\citenamefont{Spanier et~al.}(2006)}]{Spanier2006NanoLett}
\bibinfo{author}{\bibfnamefont{J.~E.} \bibnamefont{Spanier}}
  \bibnamefont{et~al.}, \bibinfo{journal}{Nano Lett.}
  \textbf{\bibinfo{volume}{6}}, \bibinfo{pages}{735} (\bibinfo{year}{2006}).

\bibitem[{\citenamefont{Sheu et~al.}(2012)}]{Sheu2012APL}
\bibinfo{author}{\bibfnamefont{Y.~M.} \bibnamefont{Sheu}} \bibnamefont{et~al.},
  \bibinfo{journal}{Appl. Phys. Lett.} \textbf{\bibinfo{volume}{100}},
  \bibinfo{pages}{242904} (\bibinfo{year}{2012}).

\bibitem[{\citenamefont{Lemanov et~al.}(1996)}]{Lemanov1996PRB}
\bibinfo{author}{\bibfnamefont{V.~V.} \bibnamefont{Lemanov}}
  \bibnamefont{et~al.}, \bibinfo{journal}{Phys. Rev. B}
  \textbf{\bibinfo{volume}{54}}, \bibinfo{pages}{3151} (\bibinfo{year}{1996}).

\bibitem[{\citenamefont{Tenne et~al.}(2004)}]{Tenne2004PRB}
\bibinfo{author}{\bibfnamefont{D.~A.} \bibnamefont{Tenne}}
  \bibnamefont{et~al.}, \bibinfo{journal}{Phys. Rev. B}
  \textbf{\bibinfo{volume}{70}}, \bibinfo{pages}{174302}
  (\bibinfo{year}{2004}).

\bibitem[{\citenamefont{Powell}(2010)}]{SHGBook}
\bibinfo{author}{\bibfnamefont{R.~C.} \bibnamefont{Powell}},
  \emph{\bibinfo{title}{Symmetry, Group Theory, and the Physical Properties of
  Crystals}} (\bibinfo{publisher}{Springer}, \bibinfo{address}{New York},
  \bibinfo{year}{2010}).

\bibitem[{Gap()}]{GapNote}
\bibinfo{note}{We take 3.6 eV as the optical band gap in our band diagram,
  based on previous publications with different Ba concentrations. The 10$\%$
  Ba concentration should have optical properties close to that of STO.}

\bibitem[{Ban()}]{BandNote}
\bibinfo{note}{The band offset is given from ref.
  \cite{Chang2002BSTOEf,Xia2011BSTOEf,Schafranek2008PRBBarrier}, which have
  different Ba doping concentrations (and thus different energy gaps energy
  gap) that are not a critical factor affecting the offset. The band offset
  depends on the sample preparation; generally, in transport measurements it
  ranges from 0.4-1.6 eV \cite{Schafranek2008PRBBarrier}. The X-ray
  photoemission data in Ref. \cite{Schafranek2008PRBBarrier} shows that the
  valence band-Fermi level separation in BSTO is 2.2-2.4 eV, indicating a band
  offset of $\sim$1.2-1.4 eV with LC(S)MO. Considering these possible variables
  (e.g., doping, gap, and different work functions), we believe our photon
  energy is greater than the band offset at the BSTO/LC(S)MO interface.}

\bibitem[{\citenamefont{Dionne}(2009)}]{MagneticOxide}
\bibinfo{author}{\bibfnamefont{G.~F.} \bibnamefont{Dionne}},
  \emph{\bibinfo{title}{Magnetic Oxides}} (\bibinfo{publisher}{Springer},
  \bibinfo{address}{New York}, \bibinfo{year}{2009}).

\bibitem[{\citenamefont{Bloch et~al.}(1996)}]{Bloch1996PRLSHG}
\bibinfo{author}{\bibfnamefont{J.}~\bibnamefont{Bloch}} \bibnamefont{et~al.},
  \bibinfo{journal}{Phys. Rev. Lett.} \textbf{\bibinfo{volume}{77}},
  \bibinfo{pages}{920} (\bibinfo{year}{1996}).

\bibitem[{\citenamefont{Mihaychuk et~al.}(1995)}]{Mihaychuk1995OLSHG}
\bibinfo{author}{\bibfnamefont{J.~G.} \bibnamefont{Mihaychuk}}
  \bibnamefont{et~al.}, \bibinfo{journal}{Opt. Lett.}
  \textbf{\bibinfo{volume}{20}}, \bibinfo{pages}{2063} (\bibinfo{year}{1995}).

\bibitem[{\citenamefont{Lu et~al.}(2008)}]{Lu2008PRBSHGT}
\bibinfo{author}{\bibfnamefont{X.}~\bibnamefont{Lu}} \bibnamefont{et~al.},
  \bibinfo{journal}{Phys. Rev. B} \textbf{\bibinfo{volume}{78}},
  \bibinfo{pages}{155311} (\bibinfo{year}{2008}).

\bibitem[{\citenamefont{Tinte et~al.}(2004)}]{Tinte2004Pr}
\bibinfo{author}{\bibfnamefont{S.}~\bibnamefont{Tinte}} \bibnamefont{et~al.},
  \bibinfo{journal}{J. Phys.: Cond. Mat.} \textbf{\bibinfo{volume}{16}},
  \bibinfo{pages}{3495} (\bibinfo{year}{2004}).

\bibitem[{\citenamefont{Adikary and Chan}(2003)}]{Adikary2003Pr}
\bibinfo{author}{\bibfnamefont{S.}~\bibnamefont{Adikary}} \bibnamefont{and}
  \bibinfo{author}{\bibfnamefont{H.}~\bibnamefont{Chan}},
  \bibinfo{journal}{Thin Solid Films} \textbf{\bibinfo{volume}{424}},
  \bibinfo{pages}{70 } (\bibinfo{year}{2003}).

\bibitem[{\citenamefont{Pontes et~al.}(2001)}]{Pontes2001Pr}
\bibinfo{author}{\bibfnamefont{F.}~\bibnamefont{Pontes}} \bibnamefont{et~al.},
  \bibinfo{journal}{Thin Solid Films} \textbf{\bibinfo{volume}{386}},
  \bibinfo{pages}{91 } (\bibinfo{year}{2001}).

\bibitem[{\citenamefont{Ruckenbauer et~al.}(2004)}]{Ruckenbauer2004Pr}
\bibinfo{author}{\bibfnamefont{V.}~\bibnamefont{Ruckenbauer}}
  \bibnamefont{et~al.}, \bibinfo{journal}{Appl. Phys. A: Mat. Sc. $\&$ Proc.}
  \textbf{\bibinfo{volume}{78}}, \bibinfo{pages}{1049} (\bibinfo{year}{2004}).

\bibitem[{\citenamefont{Lee et~al.}(2012)}]{Lee2012NLPr}
\bibinfo{author}{\bibfnamefont{O.}~\bibnamefont{Lee}} \bibnamefont{et~al.},
  \bibinfo{journal}{Nano Lett.} \textbf{\bibinfo{volume}{12}},
  \bibinfo{pages}{4311} (\bibinfo{year}{2012}).

\bibitem[{\citenamefont{Cardona}(1965)}]{CardonaPR1965}
\bibinfo{author}{\bibfnamefont{M.}~\bibnamefont{Cardona}},
  \bibinfo{journal}{Phys. Rev.} \textbf{\bibinfo{volume}{140}},
  \bibinfo{pages}{A651} (\bibinfo{year}{1965}).

\bibitem[{\citenamefont{Thielsch et~al.}(1997)}]{Thielsch1997Gap}
\bibinfo{author}{\bibfnamefont{R.}~\bibnamefont{Thielsch}}
  \bibnamefont{et~al.}, \bibinfo{journal}{Thin Solid Films}
  \textbf{\bibinfo{volume}{301}}, \bibinfo{pages}{203} (\bibinfo{year}{1997}).

\bibitem[{\citenamefont{Tian et~al.}(2001)}]{Tian2001Gap}
\bibinfo{author}{\bibfnamefont{H.~Y.} \bibnamefont{Tian}} \bibnamefont{et~al.},
  \bibinfo{journal}{J. Phys.: Cond. Mat.} \textbf{\bibinfo{volume}{13}},
  \bibinfo{pages}{4065} (\bibinfo{year}{2001}).

\bibitem[{\citenamefont{Samantaray et~al.}(2004)}]{Samantaray2004Gap}
\bibinfo{author}{\bibfnamefont{C.}~\bibnamefont{Samantaray}}
  \bibnamefont{et~al.}, \bibinfo{journal}{Phys. B: Cond. Mat.}
  \textbf{\bibinfo{volume}{351}}, \bibinfo{pages}{158 } (\bibinfo{year}{2004}).

\bibitem[{\citenamefont{Pontes et~al.}(2002)}]{Pontes2002Gap}
\bibinfo{author}{\bibfnamefont{F.~M.} \bibnamefont{Pontes}}
  \bibnamefont{et~al.}, \bibinfo{journal}{J. Appl. Phys.}
  \textbf{\bibinfo{volume}{91}}, \bibinfo{pages}{5972} (\bibinfo{year}{2002}).

\bibitem[{\citenamefont{Heber}(2009)}]{Heber2009Nature}
\bibinfo{author}{\bibfnamefont{J.}~\bibnamefont{Heber}},
  \bibinfo{journal}{Nature} \textbf{\bibinfo{volume}{459}}, \bibinfo{pages}{28}
  (\bibinfo{year}{2009}).

\bibitem[{\citenamefont{Huang}(2010)}]{Huang2010NP}
\bibinfo{author}{\bibfnamefont{H.}~\bibnamefont{Huang}}, \bibinfo{journal}{Nat.
  Photon.} \textbf{\bibinfo{volume}{4}}, \bibinfo{pages}{134}
  (\bibinfo{year}{2010}).

\bibitem[{\citenamefont{Chang et~al.}(2002)}]{Chang2002BSTOEf}
\bibinfo{author}{\bibfnamefont{S.-T.} \bibnamefont{Chang}}
  \bibnamefont{et~al.}, \bibinfo{journal}{Appl. Phys. Lett.}
  \textbf{\bibinfo{volume}{80}}, \bibinfo{pages}{655} (\bibinfo{year}{2002}).

\bibitem[{\citenamefont{Xia et~al.}(2011)}]{Xia2011BSTOEf}
\bibinfo{author}{\bibfnamefont{F.~J.} \bibnamefont{Xia}} \bibnamefont{et~al.},
  \bibinfo{journal}{J. Appl. Phys.} \textbf{\bibinfo{volume}{110}},
  \bibinfo{pages}{103716} (\bibinfo{year}{2011}).

\bibitem[{\citenamefont{Schafranek et~al.}(2008)}]{Schafranek2008PRBBarrier}
\bibinfo{author}{\bibfnamefont{R.}~\bibnamefont{Schafranek}}
  \bibnamefont{et~al.}, \bibinfo{journal}{Phys. Rev. B}
  \textbf{\bibinfo{volume}{77}}, \bibinfo{pages}{195310}
  (\bibinfo{year}{2008}).

\end{thebibliography}
\end{document}